\documentclass[11pt,twoside]{article}
\usepackage{asp2010}
\usepackage{epsf,wrapfig}

\resetcounters

\begin{document}

\title{Power-law Magnetic Field Decay and Constant Core Temperatures of Magnetars, Normal and
Millisecond Pulsars}
\author{Yi Xie $^1$ and Shuang-Nan Zhang $^{1,2}$
\affil{$^1$National Astronomical Observatories, Chinese Academy Of
Sciences, Beijing 100012, China.} \affil{$^2$Key Laboratory of
Particle Astrophysics, Institute of High Energy Physics, Chinese
Academy of Sciences, Beijing 100049, China.}}

\begin{abstract}
The observed correlations, between the characteristic ages and
dipole surface magnetic field strengths of all pulsars, can be well
explained by magnetic field decay with core temperatures of
$~2\times10^{8}$~K, $\sim2\times10^{7}$~K, and $\sim10^{5}$~K, for
magnetars, normal radio pulsars, and millisecond pulsars,
respectively; assuming that their characteristic ages are about two
orders of magnitude larger than their true ages, the required core
temperatures may be reduced by about a factor of 10. The magnetic
decay follows a power-law and is dominated by the solenoidal
component of the ambipolar diffusion mode. In this model, all NSs
are assumed to have the same initial magnetic field strength, but
different core temperature which do not change as the magnetic field
decays. This suggests that the key distinguishing property between
magnetars and normal pulsars is that magnetars were born much hotter
than normal pulsars, and thus have much longer magnetic field decay
time scales, resulting in higher surface magnetic field strength
even with the same ages of normal pulsars. The above conclusion
agrees well with the observed correlations between the surface
temperatures of magnetars and other young NSs, which do not agree
with the cooling dominated evolution of neutron stars. This suggests
a possible scenario that heating, perhaps due to magnetic field
decay, balances neutron star cooling for observed pulsars.

\end{abstract}

\section{Introduction}

The strength of the surface dipole magnetic field of a neutron star may be estimated directly from the period
and period derivative, i.e. $B_{\rm d}\approx3.3\times10^{19}(P\dot{P})^{1/2}$~G, which assumes that its
rotational energy is lost entirely via magnetic dipole radiation. Then the spin-down age (characteristic age)
can also be estimated as $\tau_{\rm c}=P/2\dot{P}$. Then all pulsars can also be displayed in the $\tau_{\rm
c}-B$ diagram (Figure 1), instead of the $P-\dot{P}$ diagram. Clearly all pulsars can be divided into three
clusters in Figure 1; from top to bottom,  the three cluster are magnetars, normal pulsars, and millisecond
pulsars, respectively.

Magnetars, which are normally referred to as neutron stars with surface magnetic fields larger than the quantum
critical value, $B_{\rm cr}=m^2 c^3/e\hbar\approx 4.4\times 10^{13}$~G, have been proposed as the most promising
energy source for Anomalous X-ray Pulsars (AXPs) and the Soft Gamma-ray Repeaters (SGRs). In the $\tau_{\rm
c}-B$ diagram, magnetars are composed of a distinctive group with higher magnetic field
($\sim5\times10^{14}-2\times10^{15} $~G) and younger characteristic ages ($\sim 10^3-10^5$~yrs).  The two
classes of X-ray Pulsars, AXPs and SGRs, differ from the common accretion-powered pulsars in massive X-ray
binaries mainly in the following observational properties \citep[see][for a review]{Mereghetti_2008}: they probably have
no binary companions, but their persistent X-ray luminosity can be larger than their spin-down power; they have
periods of activities during which they emit numerous short bursts in hard X-ray/soft gamma-ray band, and the
bursts have peak luminosity up to $\sim 10^{42}~{\rm erg}~{\rm s}^{-1}$ and last for $\sim 0.01-1$~s; they have
a short hard spike, with luminosity larger than $\sim 5\times 10^{44}~{\rm erg}~{\rm s}^{-1}$, followed by a
long pulsating tail, the so-called giant flares, a very striking phenomenon observed only from SGRs.
\begin{figure}

\centering
\includegraphics[width=4.5in]{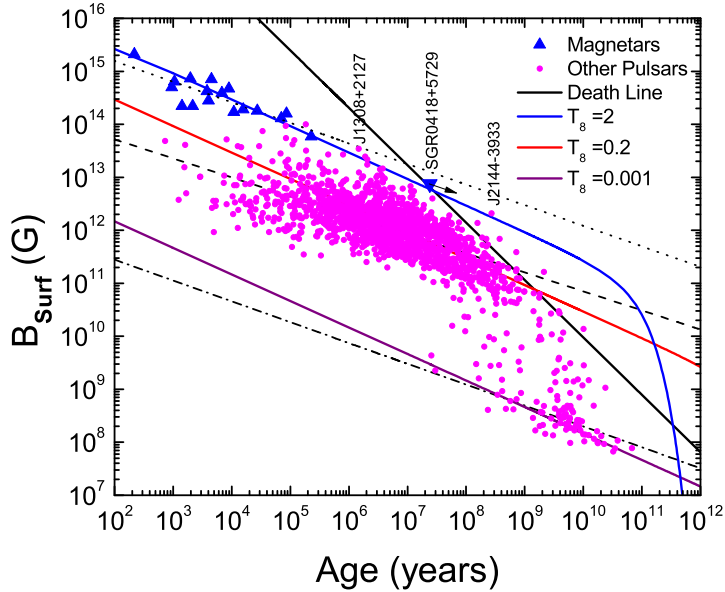}

\caption{The $\tau_{\rm c}-B$ diagram. Blue triangles: SGRs and AXPs. Magenta circles: other pulsars, including
normal radio pulsars, radio-quiet pulsars, high energy pulsars, millisecond pulsars, binary pulsars, etc. Thin
lines: the best fitted line for magnetars (dotted line), ordinary pulsars with dipole field strength
$4\times10^{10}-2\times10^{13}$~G (dashed line), and millisecond pulsars with dipole field strength $<10^9$~G
(dot-dashed line). Thick solid lines: `death line' (black), magnetic field decay for $T_8=2$ (blue), $T_8=0.2$
(red), $T_8=0.001$ (purple). For magnetars, their characteristic ages ($\tau_{\rm c}$) and surface magnetic
fields $B_{\rm surf}$ ($B_{\rm d}$) are from McGill SGR/AXP Online Catalog
(http://www.physics.mcgill.ca/~pulsar/magnetar/main.html); for other pulsars, $\tau_{\rm c}$ and $B_{\rm surf}$
are from the ATNF Pulsar Catalogue (http://www.atnf.csiro.au/people/pulsar/psrcat/).} \label{fig1}

\end{figure}

It is believed that the energy reservoir fueling the SGR/AXP activity is their extreme magnetic fields \citep[][hereafter TD96]{Thompson_1995,Thompson_1996}. For instance, their magnetic fields have enough free energy to power the
giant flares; meanwhile the short duration of their initial spikes is consistent with the propagation with
Alfv\'{e}n speed of the magnetic instability over a whole neutron star surface \citep[][]{Thompson_1995}, and
the magnetic confinement of the hot plasma is responsible for the pulsating tails \citep[][]{Mereghetti_2008}. However,
the recent discovery of a SGR with a very low surface dipolar magnetic field, SGR~0418+5729, whose dipolar
magnetic field cannot be greater than $7.5\times10^{12}$~G, implying that a high surface dipolar magnetic field
is not necessarily required for magnetar-like activities \citep[][]{Rea_2010}. And the outer gap model of magnetars
may also be challenged by the null detection in a Fermi/LAT observation of AXP 4U 0142+61 \citep[][]{Tong_2010}.
Thus, it is likely that the magnetar activity is driven by the magnetic energy stored in the internal toroidal
field \citep[][]{Thompson_1995,Thompson_2001} or the surface multipolar field. After the discovery of SGR~0418+5729, the
magnetar population includes objects with a wider range of $B$-field strengths and ages; however their
evolutionary stages still remain far from clear.

In this report, by adopting the magnetic field decay model to the magnetars, normal pulsars, and millisecond
pulsars, we show that the inner core temperatures of the three types of pulsars are systematically different,
consistent with the observed three clusters shown in Figure 1 and the observed surface temperature of magnetars
and other young normal pulsars. It is found that the surface and core temperatures of magnetars are the highest
(include the low magnetic field magnetar SGR~0418+5729) and remain constant for at least $24$~Myrs.

\section{Magnetic Field Decay of All Pulsars}

\citet[hereafter GR92]{Goldreich_1992} studied several avenues for magnetic field decay in isolated
neutron stars: ohmic decay, ambipolar diffusion, and Hall drift. Depending on the strength of the magnetic
field, each of these processes may dominate the evolution. \citet[hereafter HK98]{Heyl_1998} applied the
model to magnetars, and found that for sufficiently strong nascent fields, field decay alters significantly the
cooling evolution relative to similarly magnetized neutron stars with constant fields. Following GR92 and HK98,
a simple differential equation can be used to describe the dipole magnetic field decay (HK98)
\begin{equation}
\label{eq5}\frac{dB_{\rm p}}{dt} = -B_{\rm p}(\frac{1}{t_{\rm ohmic}}+\frac{1}{t_{\rm ambip}}+\frac{1}{t_{\rm
hall}}),
\end{equation}
where $B_{\rm p}$ is the strength of the magnetic field at the pole at the surface of a neutron star, and
$t_{\rm ohmic}$, $t_{\rm ambip}$, and $t_{\rm hall}$ are the decay time scales due to ohmic decay, ambipolar
diffusion, and Hall drift, respectively. Taking $\sigma_0=4.2\times10^{28}T_8^{-2}(\rho/\rho_{\rm nuc})^3{\rm
s}^{-1} $, where $T_8$ denotes the temperature in units of $10^8$~K, $\rho_{\rm nuc}\equiv2.8\times10^{14}~{\rm
g}~{\rm cm}^{-3}$, we have $t_{\rm ohmic}\sim2\times10^{11}\frac{L_5^2}{T_8^2}(\frac{\rho}{\rho_{\rm nuc}})^3\
{\rm yr}$, where $L_5$ is a characteristic length scale of the flux loops through the outer core in units of
$10^5$~cm. The time scales for Ambipolar diffusion are given by, $t_{\rm ambip}^{\rm
s}\sim3\times10^{9}\frac{L_5^2T_8^2}{B_{12}^2}\ {\rm yr}$, and $t_{\rm ambip}^{\rm
ir}\sim\frac{5\times10^{15}}{T_8^6B_{12}^2}(1+5\times10^{-7}L_5^2T_8^8)\ {\rm yr}$, where $B_{12}$ is the field
strength in units $10^{12}$~G, the $t_{\rm ambip}^{\rm s}$ represent for the time scale of solenoidal component
and $t_{\rm ambip}^{\rm ir}$ for irrotational component; note that the latter is only for the case of modified
URCA reactions. Finally the time scale of the Hall cascade is $t_{\rm
hall}\sim5\times10^{8}\frac{L_5^2T_8^2}{B_{12}}(\frac{\rho}{\rho_{\rm nuc}})\ {\rm yr}$.

To compare eq.~(1) with data shown in Figure 1, we take $L_5=1$, $\rho/\rho_{\rm nuc}=1$, initial field
strengths $B_{p,0}=10^{16}$~G, and begin the integrations at $t=0.1$~yr, and assume constant core temperatures
for the three classes of pulsars. In Figure 1, we plot the calculated magnetic field decay histories with
eq.~(1) for three typical core temperatures; for comparison the pulsar `death line' is also plotted \citep[][]{Chen_1993}. Several observations can be obtained from Figure 1: (i) Magnetars also follow the `death line'
(albeit with low statistics), calculated assuming the radio pulsar mechanism. (ii) The inner core temperature of
magnetars is systematically higher than other types of pulsars. The core temperature of magnetars has a very
narrow range and is about $2\times10^8$~K. The core temperatures of ordinary pulsars distribute in a relatively
wide range, from $2\times10^6$~K to $2\times10^8$~K; the center is about $2\times10^7$~K. The core temperature
of millisecond pulsars is around $10^5$~K. (iii) Several radio-quiet pulsars and normal radio pulsars fall right
on the curve for magnetars, and thus may have the same core temperature.

All observed NSs appear in the power-law parts of the model curves, i.e., $B_{\rm surf}\propto \tau_{\rm
c}^{-0.5}$. As we have shown in a companion paper in the same proceedings \citep[][]{Zhang_2011}, this corresponds
to the solenoidal component of the ambipolar diffusion dominated decay mode with a constant core temperature.
Interestingly, exactly the same relation is obtained in the magnetic dipole spin-down model with a constant $P$,
i.e., $\dot{P}\ll 1$. This immediately suggests that the observed extremely small $\dot{P}$ of NSs is due to
magnetic field decay with constant core temperature.

\section{Core Temperatures of Magnetars and Young Pulsars}
\begin{figure}
\epsscale{0.7}
\begin{center}
\includegraphics[width=5.0in]{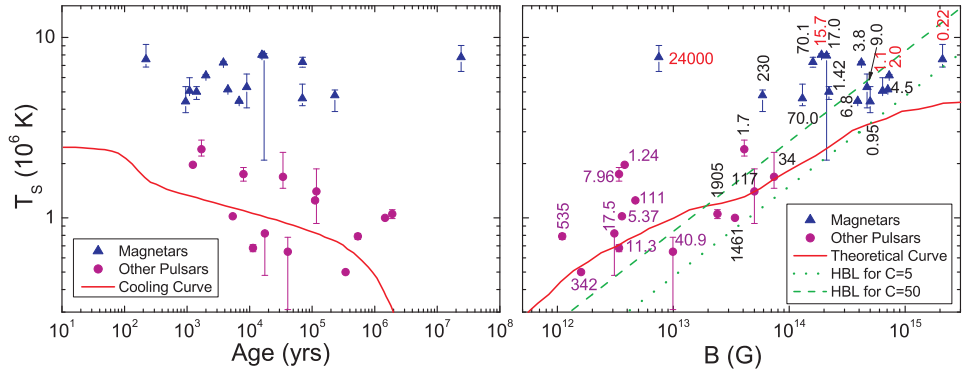}%
\end{center}

\caption{Observations of the surface temperatures of magnetars and young NSs. {\it Left panel}: surface
temperature vs. characteristic age. The red solid line is a basic theoretical cooling curve of a nonsuperfluid
neutron star with $M=1.3~M_\odot$ \citep[][]{Yakovlev_2004}. {\it Right panel}: surface temperature vs.
magnetic field. The green dotted line and dashed line are heating balance lines (HBL) for different parameters,
and the band between them is heating balance area. The red solid line is a theoretical curve for magnetothermal
evolution of magnetized neutron stars, including crustal heating by magnetic field decay (APM08). Each number
marked in the plot is the characteristic age of the pulsar in unit of $10^3$~yrs. } \label{fig2}
\end{figure}

Currently the core temperature of a NS can only be inferred from its surface temperature $T_{\rm s}$. The
spectra of some NSs are well-described by a simple blackbody (BB) radiation which is believed to be its thermal
surface emission. For NSs with spectra that include two BB components, we adopt the temperature of the component
which dominates the spectrum. In Figure 2 we show the observed $T_{\rm s}-\tau_{\rm c}$ and $T_{\rm s}-B$
relations for magnetars and young NSs. In the left panel of Figure 2, it is shown that the surface temperature
distribution can be divided into two separate groups. The normal radio pulsars and several radio-quiet pulsars
may follow the cooling curve with large scatters, though a constant temperature may also describe the data
almost equally well. However, the surface temperatures of magnetars  ($\gtrsim4\times10^6$~K) are much higher
than that of normal young NSs and also almost constant for at least twenty million years, consistent with the
model parameters used in Figure 1.

\citet{Pons_2007-AA} reported a correlation between the surface temperature and the magnetic field and
investigated similar diagram as the right panel. They approximated the correlation as $T_{\rm s,6}^4\simeq
CB_{\rm d,14}^2$, the so-called heating balance line (HBL), (see also APM08), where $C$ is a constant that
depends on the thickness of the crust with a typical value of $C\simeq 10$. They argued that the thermal
evolution of NSs with $B\gtrsim10^{13}$~G is predominantly determined by the strength of the magnetic field, and
NSs with field of $\sim10^{12}$~G should cool much more rapidly than those with field of $\sim10^{13}$~G and
higher. They further suggested that ``A NS will begin its life high on the figure with some $B_{\rm d}$. As it
cools it moves vertically downward, until decay of its field causes the trajectory to bend to the left. The star
eventually reaches its HBL, and then continues moving down it."

However, the above scenario is not supported by data; the data are more consistent with a scenario that both
magnetars and normal pulsars move horizontally to the left, but with initially very different temperatures,
i.e., their magnetic fields decay but core temperatures remain the same, again consistent with the results in
Figure 1. \citet{Rea_2010} suggested that the magnetic energy stored in the internal toroidal field that power
the violent activities for magnetars, rather than surface dipole field. \citet{Thompson_1996} investigated
the process that magnetic field decay can provide a significant source of internal heating, and the possibility
of obtaining an equilibrium between neutrino cooling and heating through magnetic field decay for
$B\sim10^{15}$~G and $T \sim10^8$~K. The internal toroidal field probably plays this role and therefore keeps
the core temperature as well as surface temperature of magnetars nearly constant.

\section{Summary and Discussion}

In this report, we found that the observed $\tau_{\rm c}-B$  correlations of all pulsars can be well explained
by magnetic field decay with core temperatures of $~2\times10^{8}$~K, $\sim2\times10^{7}$~K, and $\sim10^{5}$~K,
for magnetars, normal radio pulsars, and millisecond pulsars, respectively; the decay is dominated by the
solenoidal component of the ambipolar diffusion mode with a time scale of $t_{\rm ambip}^{\rm
s}\sim3\times10^{9}\frac{L_5^2T_8^2}{B_{12}^2}\ {\rm yr}$. With just one free parameter (the core temperature),
all data of pulsars can be reasonably well described. We emphasize that in Figure 1, all NSs are assumed to have
the same initial magnetic field strength, but different core temperature which do not change as the magnetic
field decays. This suggests immediately that the key distinguishing property between magnetars and normal
pulsars is that magnetars were born much hotter than normal pulsars, and thus have much longer magnetic field
decay time scales, resulting in higher surface magnetic field strength even with the same ages of normal
pulsars. The above conclusion agrees well with the observed correlations between the surface temperatures of
magnetars and other young NSs, which do not agree with the cooling dominated evolution of neutron stars.

In the context of temperature evolution and magnetic field decay of magnetized NSs, several sophisticated models
have developed in recent years \citep[e.g.][]{Geppert_2006,Pons_2007-AA,Aguilera_2008-AA}. In these
studies an isothermal core and a magnetize envelope were generally assumed. They showed that energy released by
magnetic field decay and Joule heating in the crust is important for the thermal evolution of a NS with field
strength $\gtrsim10^{13}$~G. Though the surface temperature of radio-quiet pulsars and normal radio pulsars may
be broadly covered by the cooling curves of different strengthes of initial magnetic field, the surface
temperature of magnetars are still too high to reach even the initial field strength is as strong as
$\thicksim10^{16}$~G \citep[][hereafter APM08]{Aguilera_2008-APJ}.

As pointed in our companion paper \citep[][]{Zhang_2011}, the characteristic ages of NSs are typically much longer than
the ages of their associated supernova remnants; the magnetic field decay is found to be responsible for this
apparent discrepancy. Using the true ages of NSs, instead of the characteristic ages, will result in reduced
core temperatures for them, by roughly one order of magnitudes; however all other conclusions reached above
remain unchanged.

The locations of all magnetars on the left side of the `death line' in Figure 1, in the same way as all other
normal pulsars, would suggest that their activities are at least related to the mechanism responsible the
radiation of normal pulsars. This is not expected in the currently accepted model for magnetars, since when
their surface magnetic fields $B$ exceeds $0.1B_{\rm cr}$, their pulsar winds may be dominated by bound pairs
rather than by freely streaming electrons and positrons \citep[][]{Usov_1996}.

 \acknowledgements  We appreciate discussions with and
comments from E. P. J. van den Heuvel, S. Tsuruta, J. L. Han, W. W. Tian, X. P. Zheng, R. X. Xu, F. J. Lu and H.
Tong. SNZ acknowledges partial funding support by the National Natural Science Foundation of China under grant
Nos. 10821061, 10733010, 10725313, and by 973 Program of China under grant 2009CB824800.

\end{document}